\documentclass[pdflatex,sn-mathphys-num]{sn-jnl}

\usepackage{graphicx}%
\usepackage{multirow}%
\usepackage{amsmath,amssymb,amsfonts}%
\usepackage{amsthm}%
\usepackage{mathrsfs}%
\usepackage[title]{appendix}%
\usepackage{xcolor}%
\usepackage{textcomp}%
\usepackage{manyfoot}%
\usepackage{booktabs}%
\usepackage{algorithm}%
\usepackage{algorithmicx}%
\usepackage{algpseudocode}%
\usepackage{listings}%

\usepackage[table]{xcolor} 
\usepackage{subcaption}
\usepackage[version=4]{mhchem}
\usepackage{tikz}
\usetikzlibrary{quantikz2}

\theoremstyle{thmstyleone}%
\theoremstyle{thmstyletwo}%

\theoremstyle{thmstylethree}%

\begin{document}

\title[Article Title]{NN-AE-VQE: Neural network parameter prediction on autoencoded variational quantum eigensolvers}


\author*[1,2]{\fnm{Koen J.} \sur{Mesman}}\email{k.j.mesman@tudelft.nl}

\author*[1, 2]{\fnm{Yinglu} \sur{Tang}}\email{y.tang-5@tudelft.nl}

\author[1,4]{\fnm{Matthias} \sur{M\"{o}ller}}

\author[1,2]{\fnm{Boyang} \sur{Chen}}

\author[1,3]{\fnm{Sebastian} \sur{Feld}}%

\affil*[1]{\orgdiv{QAIMS}, \orgname{Delft University of Technology}, \orgaddress{\street{Kluyvertweg 1}, \city{Delft}, \postcode{2629 HS}, \state{Zuid-Holland}, \country{Netherlands}}}

\affil[2]{\orgdiv{Aerospace Engineering}, \orgname{Delft University of Technology}, \orgaddress{\street{Kluyvertweg 1}, \city{Delft}, \postcode{2629 HS}, \state{Zuid-Holland}, \country{Netherlands}}}

\affil[3]{\orgdiv{Quantum Computing and Engineering}, \orgname{Delft University of Technology}, \orgaddress{\street{Mekelweg 4}, \city{Delft}, \postcode{2628 CD}, \state{Zuid-Holland}, \country{Netherlands}}}

\affil[4]{\orgdiv{Applied Mathematics}, \orgname{Delft University of Technology}, \orgaddress{\street{Mekelweg 4}, \city{Delft}, \postcode{2628 CD}, \state{Zuid-Holland}, \country{Netherlands}}}

\abstract{
Accurate simulation of many-body quantum systems remains a fundamental computational challenge. Quantum computing offers a promising path by encoding quantum mechanical systems more efficiently than classical methods. In this work, we present NN-AE-VQE, a hybrid quantum–classical framework that integrates a quantum auto-encoder (QAE) with neural network parameter prediction to reduce the resource requirements of variational quantum eigensolvers (VQE). By compressing the state space with a QAE and predicting ansatz parameters with a classical neural network, we minimize circuit depth and classical optimization overhead. We demonstrate this method on \ce{H_2} and \ce{LiH} molecules, achieving energy errors well below chemical accuracy while reducing circuit length by up to 98.3\% for \ce{LiH} simulations compared to conventional UCCSD (Unitary Coupled Cluster Singles and Doubles excitations) VQE. In doing so, we uncover critical challenges in structuring parameterized quantum circuits for machine learning, providing insight into key open questions in quantum machine learning. These results illustrate a novel approach to resource-efficient quantum simulations, enabling chemically accurate modeling of increasingly complex molecular systems on near-term quantum hardware.}




\maketitle

\section{Introduction}
\label{sec:intro}

Novel materials are crucial for improving high-impact applications such as energy storage and aerospace thermal protection systems. To reveal the atomistic origins of macroscale properties like melting point or tensile strength of a material composition, a molecular dynamics (MD) simulation is performed. However, this can prove to be a challenge as the nature of these material characteristics can be complex, for example in high entropy ceramics (HEC) \cite{zhang2019review, oses2020high}. To determine the characteristics of such complex materials, the simulations need to capture the quantum mechanical interactions accurately. It is therefore essential that the interaction between atoms (inter-atomic potential) is calculated precisely and efficiently \cite{zuo2020performance}.

However, achieving the desired accuracy at the MD scale is a considerable challenge. MD simulations are generally performed for sample size and duration in the order of $\mu m$ and $\mu s$ \cite{gilbert2021perspectives}. As such, when calculating the interaction of thousands of atoms over a significant amount of time, the computation time of these calculations becomes a serious bottleneck. While efficient methods exist, such as force fields ($O(N)$ for N electrons) and DFT ($O(N^3)$), these are too inaccurate for the e.g. the simulation of HEC. Highly accurate methods such as Unitary Coupled Cluster Single and Doubles (UCCSD)\cite{bartlett2007coupled} are computationally too costly $O(n^7)$ \cite{anstine2023longrange} limiting calculations to only a few atoms. This makes accurate simulations on a molecular dynamics scale (i.e., in the order of $10^5$ electrons) intractable.

\begin{figure*}[h]
    \centering   
    \includegraphics[width=0.9\textwidth]{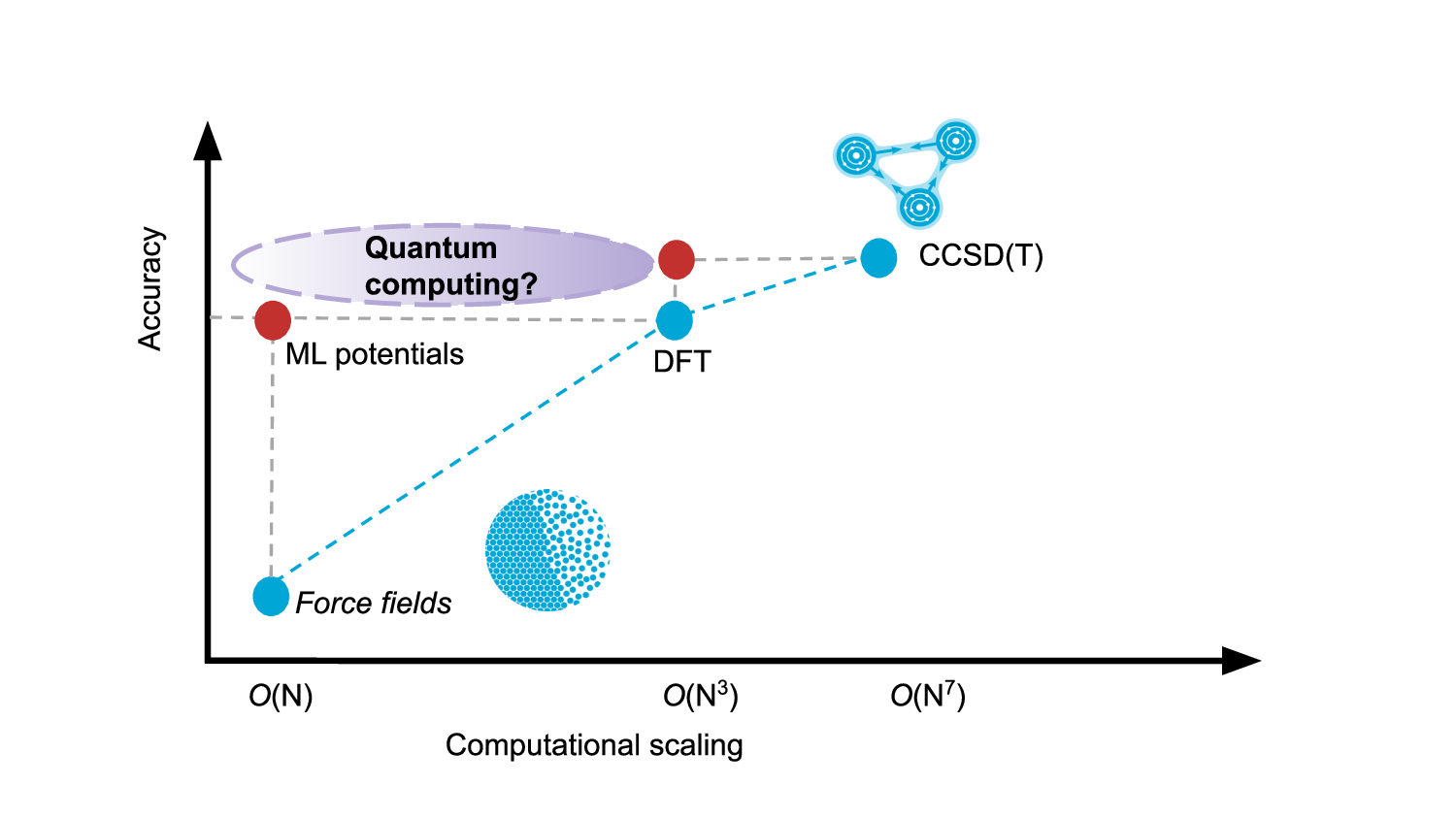}
    \caption{Computational complexity and accuracy trade-off comparison of direct computations (Force fields, DFT, CCSD, machine learning potentials). The computational efficiency of force fields to DFT allow for MD simulations, while the order of CCSD(T) computations are limited to several atoms. Recent work in ML potentials has improved upon this tradeoff, with different models having different strengths (red dots). The goal of using quantum computing is to improve this trade-off, but this has yet to be reached (purple).}
    \label{fig:scaling}
\end{figure*}

To improve the runtime/accuracy trade-off, data-based machine learning models of inter-atomic potentials have been developed \cite{pinheiro2021choosing, mishin2021machine, zuo2020performance}. These models rely on high-fidelity calculations or experimental data to train e.g. a neural network to predict the inter-atomic potential. Machine learning potentials have been demonstrated to significantly reduce the computational cost compared to conventional calculations (see also Figure \ref{fig:scaling})\cite{zuo2020performance}. However, challenges here are prevalent regarding model transferability or training data requirements \cite{anstine2023longrange, mishin2021machine, zuo2020performance}. Especially when considering long-range atomic interaction, machine learning potentials have limited performance \cite{anstine2023longrange}. Increasing the range of interactions, therefore increasing the number of atoms in the calculation, again makes the simulation intractable.

As such, we need a different computing paradigm to speed up calculations without losing accuracy. A long-theorized approach to simulating atomic systems with these benefits is quantum computing, which allows for a more efficient and direct encoding of quantum mechanical information \cite{Poulin_2009, lee2023evaluating}. In contrast to classical computers, where representing quantum states is highly inefficient due to exponential scaling, quantum computers can represent such states naturally. This leads to more accurate calculations, as encoding more system information directly correlates with higher precision.
Furthermore, quantum computing fundamentally computes entanglement (many-body interaction) more efficiently than classical methods. In doing so, larger models can be simulated more with high precision without lengthy computational costs.

Currently, quantum computers lack the required resources to effectively compute relevant inter-atomic interactions. Serious challenges exist with regard to qubit decoherence, noise, and the number of qubits. As such, the current state of quantum computers is commonly referred to as Noisy Intermediate Scale Quantum (NISQ). These challenges limit the capabilities in terms of capacity, as the number of qubits that relates to the number of molecular orbitals that can be simulated. Currently, the largest systems feature on the order of $1000$ qubits. Moreover, the circuit depth scaling with respect to system size is also a serious limitation, as recent publications have demonstrated the effort needed to simulate \ce{C2H4} (28 qubits) \cite{Cao_2022}. This indicates that in order to simulate larger systems, enough qubits are needed, and quantum circuits (ansatz) must be reduced in length to minimize the impact of hardware noise. While the development of quantum hardware's size and accuracy is steadily progressing, high-fidelity calculations on large quantum systems are not realistic in the near term. To overcome these challenges and realize high-fidelity quantum chemistry calculations on near-term hardware, quantum algorithms need to make more efficient use of the available resources by minizing qubit usage and reducing circuit length.

Most commonly, the approach to compute the ground state of a system is the variational quantum eigensolver (VQE) \cite{Peruzzo2014, hu2022benchmarking}. In variational algorithms, a parameterized quantum ansatz is evaluated for a cost function. The values of the parameters are updated through a classical optimizer until the cost function converges. This feedback loop requires computationally expensive classical-to-quantum feedback, which significantly impedes the efficiency of this algorithm. To this end, various approaches have been proposed to make the classical optimizer more efficient \cite{Sweke2020stochasticgradient}.

\subsection{Related works}

Several recent efforts have aimed to reduce the resource demands of variational quantum eigensolvers by targeting different bottlenecks, including parameter optimization, state representation, and ansatz depth.

Neural-network-assisted VQE (NN-VQE) learns to predict circuit parameters directly from problem descriptors, thereby mitigating the costly classical optimization loop and improving wall-clock efficiency \cite{Huang2023Learning, verdon2019quantum, tao2022VQEPES, miao2024neural}. While effective on small systems, these approaches typically rely on relatively deep or chemically motivated ansätze (e.g., UCC-type or multi-repetition hardware-efficient circuits). As system size grows, the depth and parameter count introduce trainability challenges (e.g., barren plateaus) and increased shot and noise sensitivity, which can limit scalability on near-term devices.

Machine learning has also been integrated with quantum computing to different ends. To combat the limitations of the small number of qubits available, quantum circuits have been developed to compress quantum information, specifically, the quantum auto-encoder (QAE) \cite{Romero_2017, anand2022quantumcompressionclassicallysimulatable, lamata2018quantum, bravo2021quantum}. The QAE is a trainable parameterized quantum circuit (PQC), which reduces the number of qubits required to store quantum information. The QAE is a quantum analog of the variational auto-encoder \cite{rumelhart1986learning}. These works, however, mainly focus on a machine learning context, and do not discuss the implementation to the VQE variational workflow.

 ADAPT-VQE builds the ansatz iteratively from an operator pool, adding gates only when they yield measurable improvement, and has been shown to reduce circuit length relative to monolithic UCCSD constructions while maintaining accuracy \cite{tang2021qubit}. However, ADAPT-VQE introduces nontrivial measurement overhead (e.g., repeated gradient or commutator evaluations over large operator pools) and its performance can be sensitive to pool design and ordering heuristics. In practice, these factors can offset depth savings with increased sampling cost, and the final circuits may still be deep for strongly correlated systems.
 
\subsection{Our work}
In this work, we address VQE’s resource demands by leveraging the QAE for state-space compression. Classical auto-encoders have not demonstrated effective compression in this setting, primarily because they cannot efficiently compress the system Hamiltonian. Instead we resort to QAE to reduce the state space. We apply the VQE algorithm to this compressed space, significantly lowering the resource requirements of VQE circuits.

\begin{figure*}[h]
    \centering
    \includegraphics[width=\linewidth]{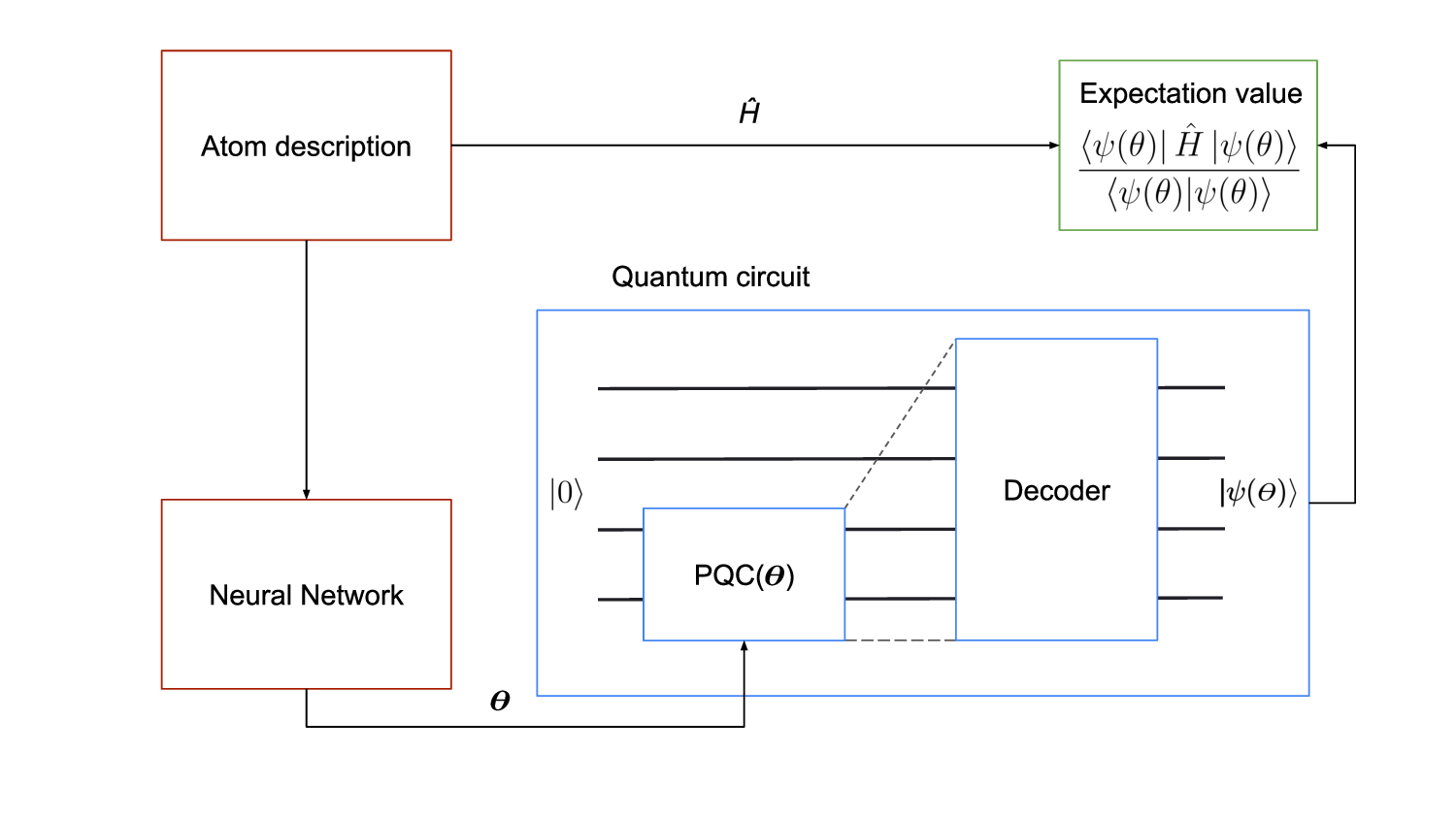}
    \caption{Overview of the VQE compression scheme. As in VQE, the atom description is used for the Hamiltonian to evaluate the expectation value. Instead of simply using a PQC on the full state space, a PQC is applied to a reduced state space to be decoded to the full state space for evaluation. While VQE generally relies on a classical optimizer to find the optimal parameter values of the PQC, we instead train a neural network to predict these values, omitting this costly subroutine.}
    \label{fig:overview}
\end{figure*}

In this work, we aim to optimize the VQE algorithm through a combination of QAE to reduce the number of qubits and apply a PQC as a hardware-efficient VQE ansatz on the latent quantum space. A classical neural network will be trained to predict the parameters of the PQC. For conciseness, we abbreviate this hybrid adaptation of VQE as NN-AE-VQE. The structure is schematically shown in Figure \ref{fig:overview}. 
This novel integration of VQE and quantum auto-encoding aims to:
\begin{enumerate}
    \item Reduce the overall VQE circuit length;
    \item Reduce the number of circuit parameters to be trained;
    \item Achieve better algorithm scaling by reducing the VQE ansatz width (number of qubits);
    \item Demonstrate the limitations of current quantum ansatz design for quantum machine learning (QML).
\end{enumerate}
                                  
We propose a QAE integration framework aimed at reducing the resource requirements of variational quantum algorithms, with applicability beyond VQE. This alleviates the current limitations of NISQ quantum computers, aiming to utilize the efficiency of quantum computing in practical applications in the near term. 

The objective of the resource reduction is to achieve better scaling. Consequently, as the resource requirements are reduced, we expect to simulate larger systems as demonstrated with unoptimized VQE simulations. To this end, we explore training the NN-AE-VQE for larger systems, such as the LiH molecule. Here, we highlight the challenges apparent in designing the PQC for quantum machine learning and its effect on training. Through these limitations, we identify key theoretical gaps needed to advance quantum machine learning.

The NN-AE-VQE method introduced is detailed in Section \ref{sec:method}. The proposed model is tested on an \ce{H_2} and \ce{LiH} molecule problem as a proof of concept and compared to baseline VQE implementations. These results are laid out in Section \ref{sec:results}. From these results, advantages, limitations, and future are discussed in Section \ref{sec:outlook}. Finally, the paper is concluded in Section \ref{sec:conclusion}.

\section{Method} 
\label{sec:method}
In this paper, we use a variational algorithm to calculate the eigenvalues of a system Hamiltonian (VQE), adapted to be applied on a compressed quantum state. The motivation for compressing a variational algorithm is to limit the algorithm depth and to reduce the number of parameters, with the added benefit of a temporary reduction of the required number of qubits.
This compression will be done through a Quantum Auto-Encoder (QAE) \cite{Romero_2017}. Furthermore, we train a classical neural network to predict the parameters of the compressed PQC. This way, we shift the computationally expensive parameter optimization to an offline trained model. Several intermediate steps are taken to integrate all elements. This includes training a quantum autoencoder, constructing a highly expressible but trainable parameterized quantum circuit (PQC), and setting up and training a classical neural network for the parameters. These steps will be explained in further detail to be combined in the final model.

\subsection{Quantum Auto-Encoding}

\begin{figure}
    \centering
    \scalebox{0.8}{
    \begin{quantikz}
        \lstick[2,brackets=none]{$\ket{\psi}$} &\gate[2][1.2cm]{\mathcal{E}}&\meter{}  &\wireoverride{n}   {\ket{0}}&\gate[2][1.2cm]{\mathcal{D}}&\rstick[2, brackets=none]{$\ket{\psi'}$}\\
        &&&&&
    \end{quantikz}
    }
    \caption{Basic structure of a quantum autoencoder (QAE). A quantum state $|\psi\rangle$ is encoded (E), after which the trash (upper) state is traced away, leaving only the encoded state (called latent space). This encoded state is decoded back into its original state space using the decoder gate $D$, of which the top (trash) input state is initialized with $\ket{0}$.\cite{Romero_2017}}
    \label{fig:QAE}
\end{figure}
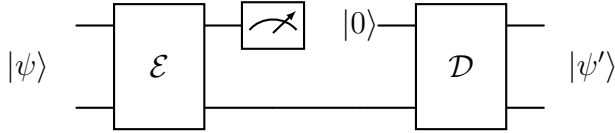

The compression of the variable quantum circuit will be done through QAE. With QAE, similar to classical auto-encoding, information is compressed to a dense form and can be decompressed to its original space. QAE aims to reduce the number of qubits to represent the information while minimizing the information loss. In this work, we will use the method presented by Romero et al. \cite{Romero_2017} since, to the best of our knowledge, it is the first implementation of QAE and offers a verifiable baseline for our results. The overall structure is presented in Figure \ref{fig:QAE}. With a QAE, a parameterized encoder is used to compress an initial quantum state to fewer qubits, which can be recovered to the original state using the decoder; the inverse of the encoder. The encoder is constructed from a parameterized quantum circuit, trained using a set of representable initial states. For these instances, a classical optimization is used to find the best parameter set for minimal information loss. In this paper, the initial state is the solved ground state of a Hydrogen molecule for a given atomic distance. Solved states are used to ensure proper coverage of the target state space. The state is prepared through a UCCSD VQE ansatz with pre-trained parameters. After encoding the state, the ``trash'' state,  i.e., the qubits discarded after encoding, should always produce a predefined state. For convenience, we use the $\ket{0}$ state. This $\ket{0}$ state is re-instantiated to decode the compressed state back to the initial state. As can be seen in Figure \ref{fig:QAE}, if the ``trash'' state equates to exactly $\ket{0}$, and the decoder is equal to the inverse of the encoder, both stages should cancel each other out resulting in matching initial and final quantum state (i.e., the fidelity $F(\ket{\psi_i}, \rho_i^{out}) = 1$). To train the decoder, it suffices to use only the ``trash'' state \cite{Romero_2017}. This method is also employed in this work, using a statevector simulator. To evaluate the ``trash'' state, the qubits in latent space are traced away, and the fidelity of the ``trash'' state is extracted. On physical quantum hardware, a SWAP test \cite{barenco1997stabilization} should be used to compare the ``trash'' state to a reference state $\ket{0}$.

Moreover, several different quantum states must be used in training to ensure good accuracy over a range of quantum states. In this work, we use six random states to train the circuit, similar to the work of Romero et al.

\subsection{PQC in latent space} \label{sec:training}
\begin{figure}[h]
    \centering
    \scalebox{0.65}{
    \begin{quantikz}
        &\gate{U_{3}}&\ctrl{1}&&&\targ{}&\gate{U_{3}}&\ctrl{2}&&\targ{}&&\\
        &\gate{U_{3}}&\targ{}&\ctrl{1}&&&\gate{U_{3}}&&\ctrl{2}&&\targ{}&\\
        &\gate{U_{3}}&&\targ{}&\ctrl{1}&&\gate{U_{3}}&\targ{}&&\ctrl{-2}&&\\
        &\gate{U_{3}}&&&\targ{}&\ctrl{-3}&\gate{U_{3}}&&\targ{}&&\ctrl{-2}&
    \end{quantikz}
    }
    \caption{Circuit of the fully entangled PQC for 4 qubits.}
    \label{fig:ansatz}
\end{figure}
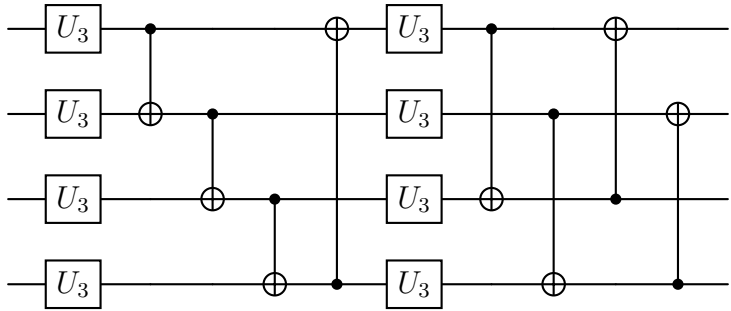

In this work, we explore the usage of applying an ansatz in the latent space. With a pre-trained QAE, a parameterized quantum circuit (PQC) can be applied to the encoded quantum state. Several approaches can be taken to construct the PQC, such as a hardware-efficient modular circuit \cite{Kandala_2017}, or an application-specific adaptive PQC \cite{ferrari2022adaptive}, and others \cite{hu2022benchmarking}. For our approach, a more generic PQC is required, as the structure of the encoded quantum data is not fully known as it is determined by the QAE parameters. In this work, a pre-defined structure, specifically the \textit{strongly entangled} ansatz \cite{Schuld_2020}, is used as an expressive, but heuristic (generic) PQC. This PQC is shown in Figure \ref{fig:ansatz}. While this PQC ensures high expressivity, which is especially relevant for an encoded state space, it could be further optimized to reduce circuit depth.

In the used ansatz, the gate $U_3$, a generic 3-angle ($\theta$, $\pi$, $\gamma$) single qubit gate, is expressed as \cite{openqasm}:

\begin{equation}
    U_3 = \begin{bmatrix}
        \cos(\frac{\theta}{2}) & -e^{i\lambda}\sin(\frac{\theta}{2})\\
        e^{i\phi}\sin(\frac{\theta}{2})&e^{i(\phi+\lambda)\cos(\theta/2)}
    \end{bmatrix}
\end{equation}

For $n$ qubits, the ansatz uses $2n$ single qubit gates ($U_3$) and $2n$ 2-qubit gates (CNOT), with $6n$ parameters.

\subsection{Training the AE-VQE}
Using a classical neural network to predict VQE parameters has been demonstrated before, e.g., by Tao et al. \cite{tao2022VQEPES}. Here, a UUCSD ansatz is used for VQE \cite{Peruzzo2014, romero2018strategies}, and a simple neural network is trained to predict the parameters.

The training of the PQC in combination with the QAE is shown schematically in Figure \ref{fig:dec-vqe}. First, the PQC is applied to a reduced $\ket{0}$ state to create the ground state in latent space, followed by applying the pre-trained decoder (inverse encoder) for evaluation. The circuit evaluation is done through state estimation for the molecular Hamiltonian. Obtaining the training data will be expanded upon in Section \ref{sec:params}.

\begin{figure*}[h]
    \centering
    \begin{quantikz}
        &\wireoverride{n} \lstick[2]{$\ket{0}^{\otimes n-k}$}&&\gate[4]{\mathcal{D}}
        \hphantom{wide}&\rstick[4, brackets=none]{$\ket{\psi_{\theta}}$}\\
        &\wireoverride{n}&&&\\
        &\wireoverride{n} \lstick[2]{$\ket{0}^{\otimes k}$}&\gate[2]{PQC(\theta)}&&\\
        &\wireoverride{n}&&&\\
        \lstick{System \\description}&\gate[2]{NN} \setwiretype{c} &\wire[u]{c} \rstick{$ \overrightarrow{\mathbf{\theta}} $}
    \end{quantikz}

    \caption{Using a classical neural network to predict compressed ansatz parameters. A classical optimizer finds optimal values for the ansatz with a pre-trained decoder circuit. The found parameters are then used as training data for the classical neural network.}
    \label{fig:dec-vqe}
\end{figure*}
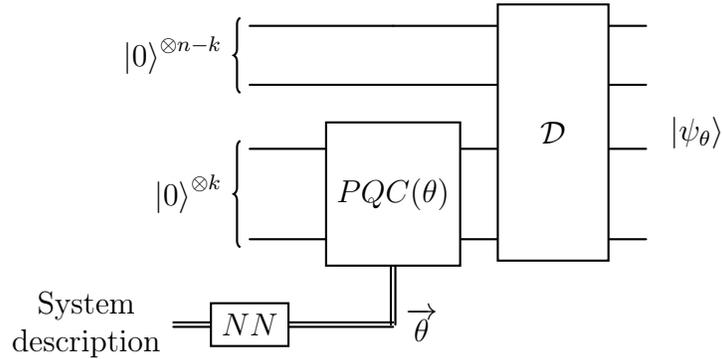

\begin{figure*}[h]
    \centering
    \begin{subfigure}{0.45\textwidth}
        \centering
        \begin{quantikz}
            &\gate[2]{VQE}&\gate[2]{PQC^{\dagger}(\theta)}&\meter[2]{}\\
            &&&
        \end{quantikz}

        \caption{Direct comparative measurement.}
        \label{subfig:direct}

    \end{subfigure}
    ~
    \begin{subfigure}{0.45\textwidth}
        \centering
        \begin{quantikz}
            &\gate[]{H}&\targ{}&\targ{}&\gate[]{H}&\meter{}\\
            &\gate[2]{VQE}&\ctrl{-1}&&&\\
            &&&\ctrl{-2}&&\\
            &\gate[2]{PQC(\theta)}&\ctrl{-2}&&&\\
            &&&\ctrl{-2}&&
        \end{quantikz}
        \caption{Measurement through SWAP-test.}
        \label{subfig:SWAP}

    \end{subfigure}
    ~
    \begin{subfigure}{0.7\textwidth}
        \centering
        \begin{quantikz}
            &\gate[2]{VQE}&\meter[2]{} &\wireoverride{1} \rstick[4]{ $\left\lvert \frac{ \bra{\psi_{PQC(\theta)}} H \ket{\psi_{PQC(\theta)}} }{ \braket{\psi_{PQC(\theta)}}{\psi_{PQC(\theta)}} } - \frac{ \bra{\psi_{VQE}} H \ket{\psi_{VQE}} }{ \braket{\psi_{VQE}}{\psi_{VQE}} } \right\rvert$ }\\
            &&&\wireoverride{1}\\
            &\gate[2]{PQC(\theta)}&\meter[2]{}&\wireoverride{1}\\
            &&&\wireoverride{1}
        \end{quantikz}

        \caption{Direct comparison of expectation values.}
        \label{subfig:exp}

    \end{subfigure}
    
    \caption{Alternative configurations for matching a PQC to a VQE state.}
    \label{fig:datagen}
\end{figure*}
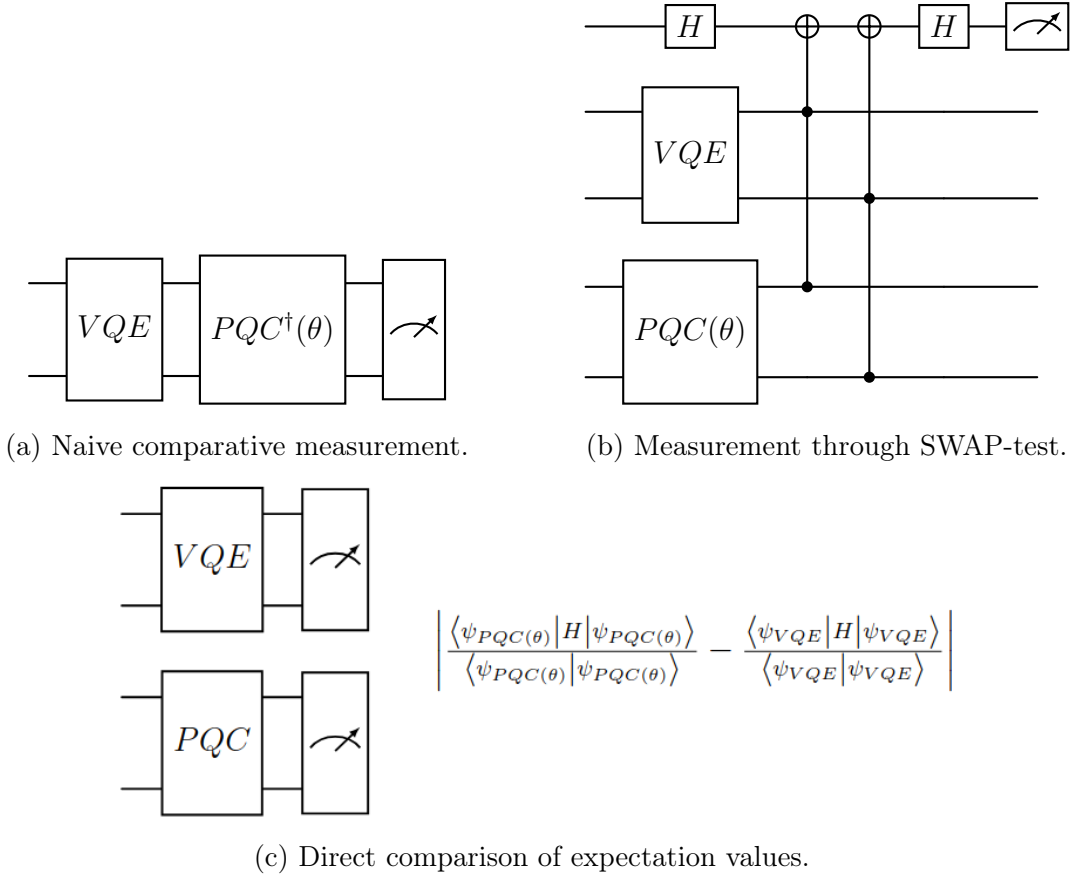

Generating the training data for the neural network is done through optimizing the PQC to match the true VQE state. The evaluation of comparing the VQE state to the PQC state can be done through several configurations. These configurations are depicted in Figure \ref{fig:datagen}. Here, the state space is depicted as 2 qubits for simplicity, but in reality consists of the $N$ qubits of the original VQE state space. The PQC circuit is simplified, representing the PQC with decoder configuration shown in Figure \ref{fig:dec-vqe}.

Three methods were considered for finding the optimal parameters $\theta$ for the trainable PQC.
The initial approach was to directly compare the pre-trained VQE circuit with the inverse of the trainable PQC, as shown in Figure \ref{subfig:direct}. Here, the VQE circuit is followed by the inverse of a unitary $U(\theta)$ where the parameters are trained such that $U(\theta)$ equals the VQE circuit. If the VQE ansatz and $PQC^{\dagger}(\theta)$ are equal, one could naively expect to measure only $\ket{0}$, as both gates would cancel. While this is a viable approach, this configuration makes for a deep quantum circuit, making training slow.
The second approach is through a SWAP-test, as depicted in Figure \ref{subfig:SWAP}. This method significantly reduces the circuit depth: the VQE and PQC are placed in parallel, with small overhead for the swap test. Furthermore, this method only needs a single qubit measurement in the computational basis. The drawback of this method is the $2N+1$ qubit requirement for an $N$-qubit system state. As a quantum simulator is used, the doubling of qubits vastly increases the runtime. 
In the final method, the expectation value of the trainable PQC is compared to the (validation) expectation value of the VQE ansatz (see Figure \ref{subfig:exp}). This approach is the simplest, which also proved to be the fastest while achieving low error rates. In this method, this means only $N$ qubits are needed at a time, with the VQE results pre-trained and stored in a database. It should be noted that while VQE is here used as an example, alternative methods of obtaining reference values can be used, such as experimental data of high-fidelity classical calculations which can be extracted from available databases.

\subsection{Parameter trainability} \label{sec:params}
\begin{figure}[h]
    \centering
        \includegraphics[width=0.6\linewidth]{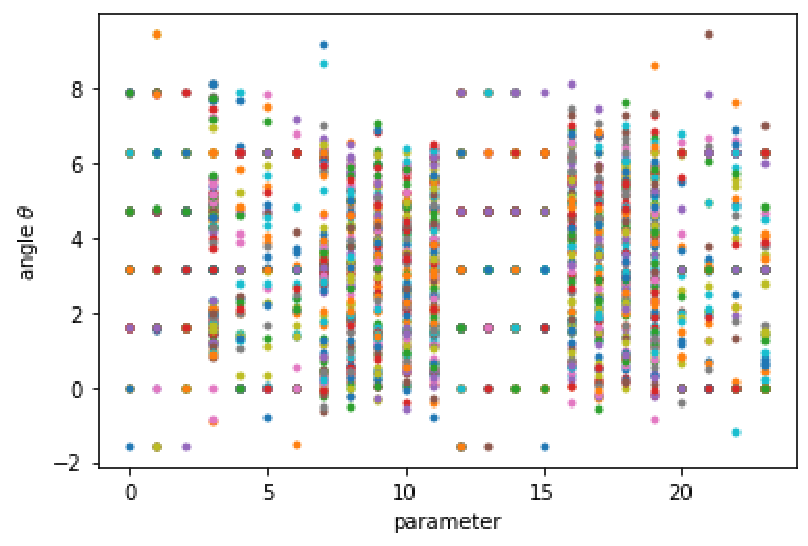}
         \caption{Parameter angles obtained using random initialization. The colors represent different samples. The discrete spread in angles for most parameters indicates that different angle combinations are found for implementing the same qubit rotations.}
        \label{subfig:uncorr}

\end{figure}

To train the classical neural network to predict the parameters, several improvements are necessary to make the parameters learnable. We chose the ansatz to be expressible to ensure coverage of the solution space, but this poses the challenge that, while the parameter optimizations converge with good accuracy, the found parameters differ largely due to the many local optima. Figure \ref{subfig:uncorr} shows several samples of these parameters, with very small overlap. While the parameters differ significantly per sample, the combination of the parameters often show to represent identical single-qubit gates. More specifically, the three parameters of the individual U3 gates find several parameter combinations for the same gate, and as a circuit, different gate combinations find a similar final state (i.e., local optima). This causes a simple neural network to average between these parameter values, resulting in inaccurate results.

\begin{figure*}[ht]
    \centering
    \begin{subfigure}[t]{0.48\linewidth}
        \includegraphics[width=\linewidth]{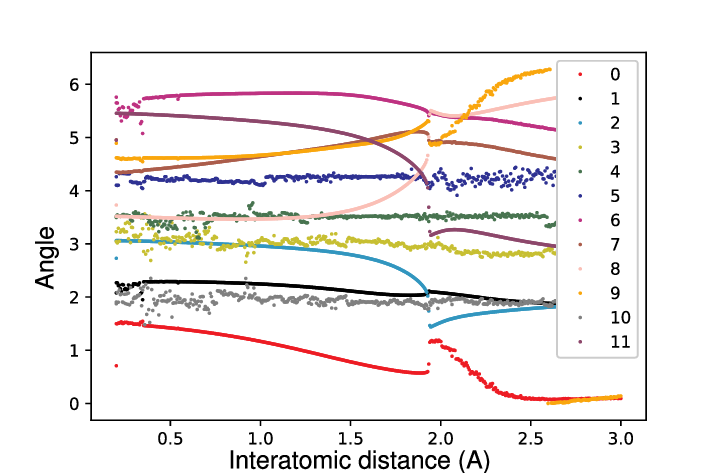}
        \caption{Parameter angles obtained using a global warm start: optimization starts from the optimized parameters of a previously found sample.}
        \label{subfig:disc}
    \end{subfigure}
    \begin{subfigure}[t]{0.48\linewidth}
        \includegraphics[width=\linewidth]{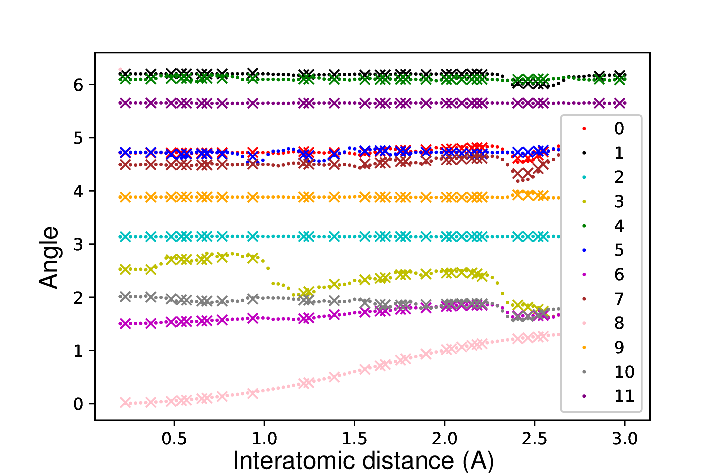}
        \caption{Parameter angles obtained using a global warm start with sequentially optimized gates. Additionally, the optimization steps are bound to ensure smoother parameter progression.}
        \label{subfig:smooth}
    \end{subfigure}

   \caption{Optimized parameters for the strongly entangled ansatz for the H-H bond, sampled from different bond distances. Each plot contains 100 samples of optimized points. The x-axis shows the bond length, with the corresponding optimized angles on the y-axis. Parameters are presented by their index (0-11).}
   \label{fig:params_2}
\end{figure*}

Using the optimized parameters of a single sample as a warm-start of the optimizations of all samples, the angle of a parameter for different samples align significantly better compared to optimization without warm-start. However, gate symmetries (different combinations of gate parameters constructing the same quantum operation) still remain a challenge. In order to train a neural network to predict the PQC parameters, we require the parameters to have a smooth progression with regard to interatomic distance. However, in Figure \ref{subfig:disc} is shown that due to gate symmetries, a discontinuity can occur. The parameters jump from one gate configuration (symmetry) to another, while still forming the same quantum operation. Training a neural network on data with such a discontinuity is significantly harder and should be avoided.

To circumvent such parameter discontinuities, the parameter search is constrained for each new atomic distance, such that for each step $t$, the parameters $\vec{\boldsymbol{\theta}}_t$, the bounds are set with tunable parameters $\alpha$ and $\beta$, and updated as:
\begin{equation} \label{eq:con}
    \vec{\boldsymbol{\theta}}_t = \vec{\boldsymbol{\theta}}_{t-1} + \delta \pm \alpha |\delta|+\beta,\quad \delta = \vec{\boldsymbol{\theta}}_{t-1} - \vec{\boldsymbol{\theta}}_{t-2}
\end{equation}

The settings that we recommend are $\alpha = \frac{10}{\text{num samples}}$, and $\beta = 0.2$ to allow some convergence in case the gradient is $0$ at any step. Using these bounds, the result in Figure \ref{subfig:smooth} no longer shows any discontinuity, allowing for better training. Still, the obtained parameters show that many parameters do not change, which could be used to simplify the ansatz. This would reduce the total number of parameters. Initial results show small reductions in accuracy, but require further work. 

\section{Results}
\label{sec:results}

In this section, we evaluate the NN-AE-VQE for accuracy and efficiency in terms of circuit depth and number of parameters.
To do so, we test against established algorithms as verifications. To demonstrate the encoder and neural network model, we used an \ce{H2} molecule as a minimal benchmark. Furthermore, we scale the system to a \ce{LiH} molecule and reflect on the encountered challenges. Due to the challenges encountered with scaling, larger molecules have not been demonstrated. For the electronic structure calculations, the error rate needs to be below chemical accuracy ($1~\text{kcal}/\text{mole} \approx 0.00159 ~\text{Hartree}$).

\subsection{Achieved QAE accuracy for \ce{H2}}
The first step in establishing the model is training the QAE. As errors propagate in each step, it is crucial to ensure the QAE implementation achieves high accuracy. We demonstrate the compression of 4 qubit \ce{H2} states to 2 qubits, using the configuration shown in Figure \ref{fig:QAE}.

\begin{table*}[]
    \centering
    \begin{tabular}{|c|c|c|}
    \hline
    Compression & Training fidelity & Validation accuracy (RMSE) \\
    \hline
    4-3   &  $3.16 \cdot 10^{-13}$ & $1.18 \cdot 10^{-13} (\pm 1.99 \cdot 10^{-13})$ \\
    4-2   & $2.22 \cdot 10^{-11}$ & $8.38 \cdot 10^{-10} (\pm 8.70 \cdot 10^{-10})$ \\
    4-1   &  $1.72 \cdot 10^{-3}$& $5.56 \cdot 10^{-4} (\pm 7.26 \cdot 10^{-4})$\\
    \hline

    \end{tabular}
    \caption{Training and validation results of the QAE, using 6 training samples for different compression rates. Validation accuracy is shown as RMSE with standard deviation.}
    \label{tab:H2_QAE}
\end{table*}

The training of the QAE for this small size required only 6 sample points to train in order to get sufficient accuracy. As for demonstration purposes, we used a statevector simulation, where the evaluations of the different samples are executed in parallel, and training can be done in a matter of minutes. The results of training the QAE are summarized in Table \ref{tab:H2_QAE}. A training fidelity loss of the trash state of $2.22 \cdot 10^{-11}$ was achieved, which measures a mean absolute error (MAE) of the \ce{H2} energy of $8.38 \cdot 10^{-10}$. The energy loss being higher compared to the training loss is expected, as the state fidelity will always be equal to or lower than the trash state fidelity \cite{Romero_2017}. As the loss from compression cannot be restored, the error of any later training result will not be lower than the compression error. For achieving chemical accuracy ($\approx 1.59 e^{-3}$,) the achieved accuracy can be considered accurate enough.

\subsection{NN-AE-VQA accuracy and scaling}
We now evaluate the PQC ansatz in combination with the decoder on the obtained training data. Using these parameters, we can evaluate the accuracy we can gain using a compressed ansatz. Furthermore, we use the results as training data for the classical neural network.

The neural network used for predicting the parameters is set up as a simple all-to-all sequential neural network, which uses the inter-atomic distance as the input value and VQA parameters as the output. The neural network is implemented with 4 hidden layers of 30 nodes each. This configuration was found to give the best results in tests. Input values are normalized and output values are translated back to a $2 \pi$ range. Additionally, to calculate the loss function the angles are translated to the squared distance of Cartesian coordinates, to ensure angle distances take rotational symmetry into account:
\begin{equation}
\begin{split}
    C = (\text{cos}(\theta_{true})-\text{cos}(\theta_{pred}))^2 \\
    + (\text{sin}(\theta_{true})-\text{sin}(\theta_{pred}))^2
\end{split}
\end{equation}
with parameter prediction $\theta_{pred}$ and actual value $\theta_{true}$.
Alternatively, the more common cosine similarity loss function can be used. However, this performed slightly worse in our findings.

\begin{figure}[h]
    \centering
    \includegraphics[width=0.6\linewidth]{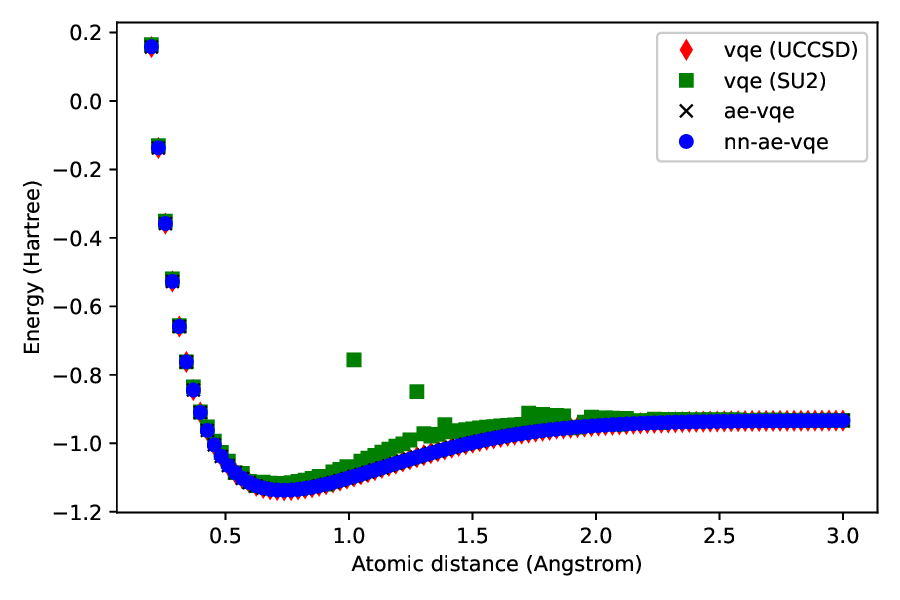}
    \caption{Comparison energy values for the $H_2$ molecule for different vqe methods.}
    \label{fig:inter-potential}
\end{figure}

\begin{figure*}[h]
    \centering
    \includegraphics[width=0.9\linewidth]{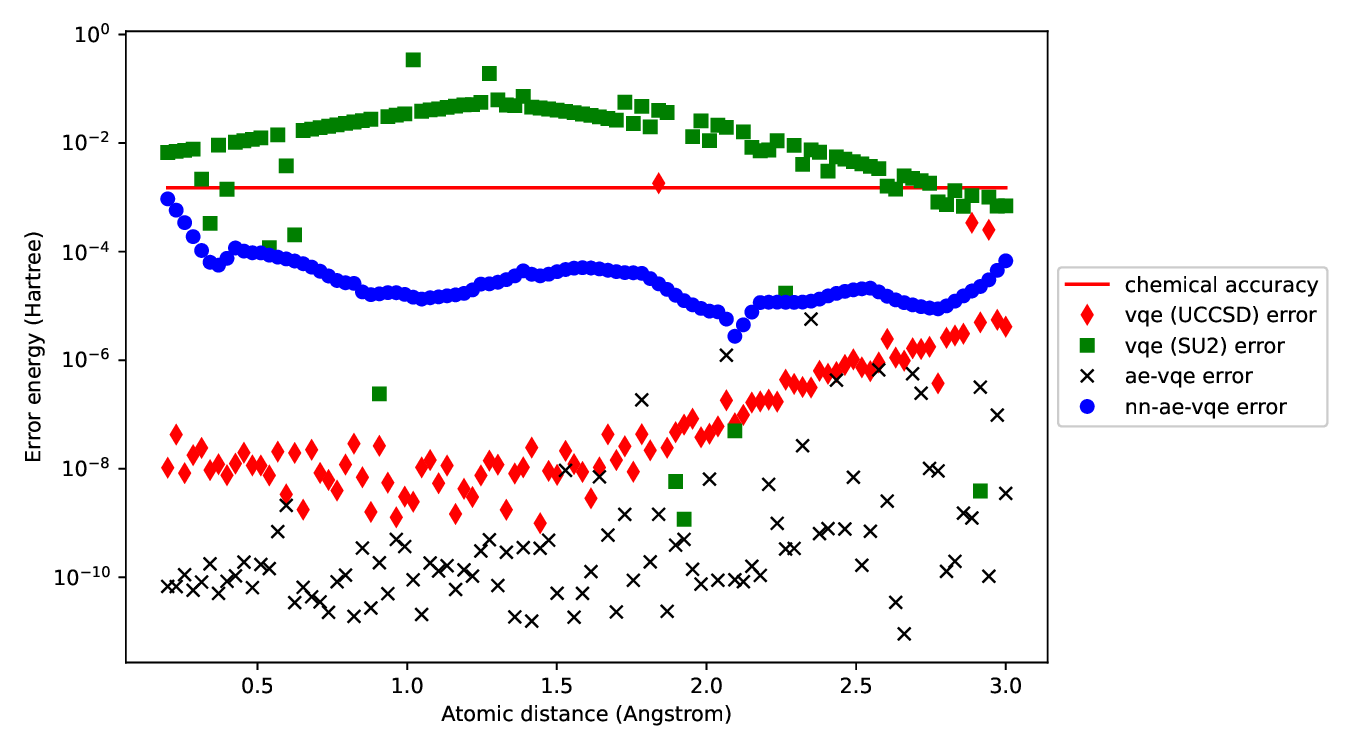}
    \caption{Comparison of errors for 100 samples. The displayed methods are UCCSD VQE, SU2 VQE, the training data for the neural network, and the neural network predictions. The neural network training data is obtained for the PQC and the pre-trained QAE, optimized for the optimal values as described in Section \ref{sec:params}.}
    \label{fig:errors}
\end{figure*}

\begin{table*}[h]
    \centering
    \begin{tabular}{|c|p{40mm}|c|p{20mm}|p{20mm}|p{20mm}|}
        \hline
         & RMSE error & \#gates & Parameters \\
        \hline
        efficient SU2 & $4.69 \cdot 10^{-2}(\pm 4.03 \cdot 10^{-2})$  & 60 & 32 \\
        UCCSD & $1.86 \cdot 10^{-4}(\pm 1.85 \cdot 10^{-4})$ & 150 & 3 \\
       AE-VQE & $5.95 \cdot 10^{-7} (\pm 5.87 \cdot 10^{-7})$&$6 + 36$ & 12 \\
        NN-AE-VQE & $ 1.23 \cdot 10^{-4} (\pm 1.12 \cdot 10^{-4})$ & $6 + 36$ & 12 \\
        \hline
    \end{tabular}
    \caption{Accuracy, gate requirements and number of parameters for different VQE implementations compared to (NN)-AE-VQE.}
    \label{tab:errors}
\end{table*}

The training data for the neural network was generated by optimizing the parameters, starting from the optimal parameters of a pre-obtained sample. The points are optimized using the warm start of the sample point, using the constraints defined by Equation \ref{eq:con}.

The results of the neural network predictions are benchmarked against the training data, the UCCSD ansatz and the SU2 ansatz. The UCCSD ansatz is known to be an accurate ansatz for VQE, but its circuit depth suffers from exponential scaling with the number of qubits. The efficient SU2 ansatz, on the other hand, is more of a heuristic, and in that, more efficient in terms of circuit depth scaling. However, hardware efficient ansatze generally require more parameters and perform worse in terms of accuracy. Both the UCCSD and SU2 ansatze are implemented directly from the Qiskit library \cite{qiskit2024}. The SU2 uses the default settings with an RY and RZ gate, and three repetitions, as this shows empirically the best results. All results are compared to exact diagonalization.

The results in Figure \ref{fig:errors} and Table \ref{tab:errors} show performance in line with expectations. The training data (black crosses) is more accurate compared to the VQE UCCSD reference results (red diamonds). Logically, the neural network predictions give rise to more errors, with the limited accuracy of the parameter predictions. Despite these errors, the neural network results are well below the target chemical accuracy. Moreover, on average the neural network is more accurate compared to the SU2 results (green squares).

Overall, the results of NN-AE-VQE look promising. Sufficient accuracy can be achieved through AE-VQE, requiring shorter quantum circuits and fewer parameters. Even by implementing a simple neural network to predict the AE-VQE parameters, an error rate below chemical accuracy can be achieved. Furthermore, the neural network predicted parameters could also serve as a warm-start for AE-VQE calculations. This opens the path to scaling this method and further refining it.

\subsection{Accuracy and scaling of NN-AE-VQE with LiH}
It is crucial to explore how the accuracy of NN-AE-VQE scales with the number of qubits to evaluate its effectiveness for physically relevant systems. To this end, the above steps have been repeated for the LiH system. Using a tapered off electron orbital and the parity mapping, the system was encoded in 8 qubits, as opposed to 12.

\begin{table*}[]
    \centering
    \begin{tabular}{|c|c|c|}
    \hline
    Compression & Training fidelity & Validation accuracy (RMSE) \\
    \hline
    8-7   &  $4.59 \cdot 10^{-12}$ & $4.07 \cdot 10^{-12} (\pm 6.98 \cdot 10^{-13})$ \\
    8-6   & $8.84 \cdot 10^{-12}$ & $1.28 \cdot 10^{-12} (\pm 1.32 \cdot 10^{-13})$ \\
    8-5   &  $6.33 \cdot 10^{-12}$& $4.18 \cdot 10^{-12} (\pm 3.59 \cdot 10^{-13})$\\
    8-4 & $5.84 \cdot 10^{-12}$ & $5.47 \cdot 10^{-12} (\pm 1.68 \cdot 10^{-13})$ \\
    8-3 & $0.11$ & $0.40 (\pm 0.35)$\\
    \hline

    \end{tabular}
    \caption{Training and validation results of the QAE, using 10 training samples for different compression rates. Validation accuracy is shown as RMSE with standard deviation.}
    \label{tab:LiH_QAE}
\end{table*}

The QAE for this system was implemented using a single layer of the hardware-efficient SU2 ansatz. It should be noted that using the ansatz proposed by Romero et al. \cite{Romero_2017}, used for the \ce{H_2} QAE, was not accurate enough for compressing the \ce{LiH} states with high accuracy ($3.42\cdot 10^{-5}$ training error, $1.65 \cdot 10^{-5}$ validation error). Training was done using 10 reference states, and a measured average training fidelity loss of $5.84 \cdot 10^{-12}$ (of the 10 training samples) was achieved for an 8 to 4 qubit compression. Validating the QAE using 50 reference states (with varying interatomic distance between [0.2, 3.0] $\mathring{A}$) resulted in an achieved accuracy of $5.47 \cdot 10^{-12}$ RMSE. The compression results for all tested compression rates are presented in Table \ref{tab:LiH_QAE}.  We note that the accuracy for 8 to 3 qubits compression drops significantly in accuracy, even for more ansatz layers. This could indicate a limit to the rate the LiH state can be compressed without significant losses. This error for a 8 to 4 compression is still well below chemical accuracy. For this reason, in the further training, the 8 to 4 compression will be used.

\begin{figure}[h]
    \centering
    \includegraphics[width=0.8\linewidth]{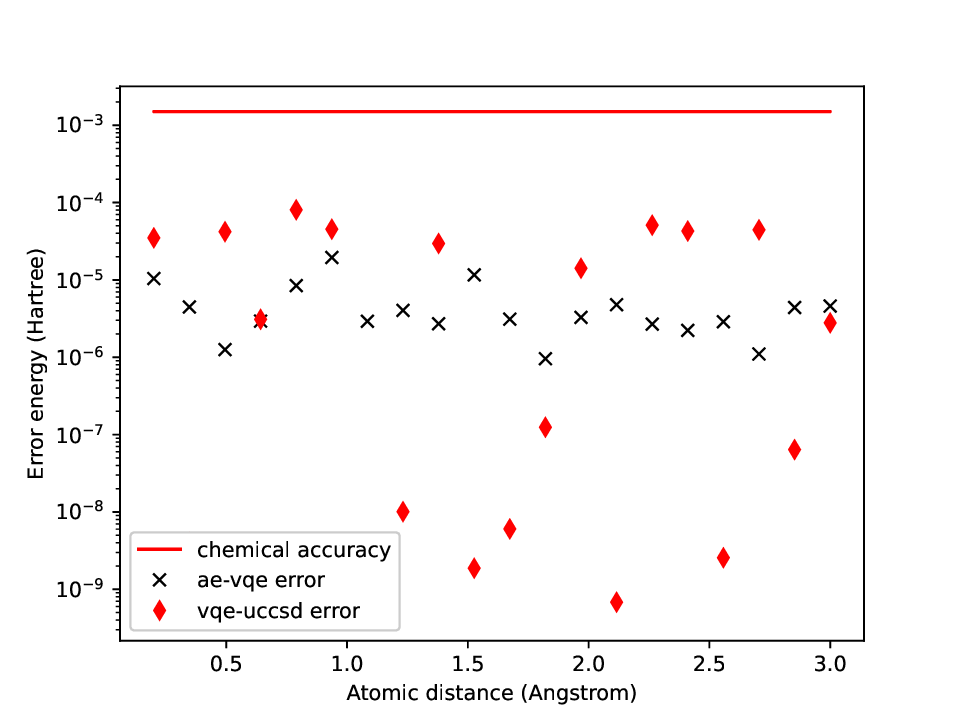}
    \caption{Validation results for the LiH molecule using 8 to 4 qubits compression. The results are obtained using a 3-layer Two local ansatz (U3 mixer, circular entanglement). Errors shown are with respect to exact diagonalization.}
    \label{fig:LiH_acc}
\end{figure}

\begin{table*}[]
    \centering
    \begin{tabular}{|c|c|c|}
    \hline
    Ansatz structure & Layers & Accuracy (RMSE) \\
    \hline

    EfficientSU2 & 3 & $1.08 \cdot 10^{-2} (\pm 1.02\cdot 10^{-2})$\\
    EfficientSU2 & 4 & $8.94\cdot 10^{-3} (\pm 8.70 \cdot 10^{-3})$\\
    \rowcolor{gray!20}
    Two local U3 circular & 3 & $1.25 \cdot 10^{-5} (\pm 8.18 \cdot 10^{-6})$\\
    Two local U3 circular & 4 & $6.57 \cdot 10^{-6} (\pm 4.36 \cdot 10^{-6})$\\
    
    Two local rxry sca & 3 & $3.06 \cdot 10^{-4} (\pm 2.79 \cdot 10^{-4})$\\
    Two local rxry sca & 4 & $2.01 \cdot 10^{-5} (\pm 1.64 \cdot 10^{-5})$\\
    \rowcolor{gray!20}
    VQE-UCCSD & 4 & $3.09 \cdot 10^{-5} (\pm 2.38 \cdot 10^{-5})$ \\
    \hline

    \end{tabular}
    \caption{Obtained AE-VQE accuracies for different ansatz configurations. Ansatz consist of 4 qubit wide circuits, applied to the 8-4 qubit QAE. Notably, the Two local ansatz with U3 mixer layers, circular entanglement and only 3 layers, performs comparably to the UCCSD-VQE ansatz with 4 layers.}
    \label{tab:LiH}
\end{table*}

For obtaining the accuracy of the AE-VQE, various hardware-efficient ansatze with different numbers of layers were explored for their effectiveness. The results have been summarized in Table \ref{tab:LiH}. The obtained results show that using hardware-efficient ansatze is eventually sufficient to reach chemical accuracy using the AE-VQE structure. The results for the 3-layer two local ansatz are shown in Figure \ref{fig:LiH_acc}.

Using the 8-4 compression, clear advantages can be seen for the AE-VQE method. Compared to using a straightforward hardware efficient ansatz on 8 qubits, applying the ansatz to only 4 qubits halves the number of parameters from 96 to 48 for 3 layers. This significantly speeds up optimization time. In terms of circuit depth, the hardware efficient ansatz scales linear (with the exception of full entanglement) and therefore does not result in significant gains here. Compared to the UCCSD ansatz on the other hand, shows much higher gains. The AE-VQE implementation requires a depth of $8\text{ (QAE)} + 17 \text{ (Two local U3 ansatz, 3 layers)}$ gates, while the UCCSD VQE requires 1636 gates. This is a reduction of $98.3 \%$ of the circuit length. In terms of accuracy, AE-VQE slightly outperforms UCCSD VQE, with an AE-VQE RMSE error of $1,25 \cdot 10^{-5} (\pm 8.18 \cdot 10^{-6})$ compared to the UCCSD RMSE error of $3.09 \cdot 10^{-5} (\pm 2.38 \cdot 10^{-5})$. Both results are presented in Figure \ref{fig:LiH_acc}.

Due to the challenge of finding an ansatz capable of producing chemically accurate results, the experiments have not been scaled further. These results shown, however, do give promise of being capable of solving larger molecules.

\section{Discussion and outlook}
\label{sec:outlook}

In this section, we will discuss our observations in training and validating the NN-AE-VQE algorithm, and we will discuss potential methods that could be explored to improve upon the proposed method further. 

\subsection{Quantum circuit scheduling}
\begin{figure*}
       \centering
        \includegraphics[trim={0 5cm 0cm 4cm},clip, width=\linewidth]{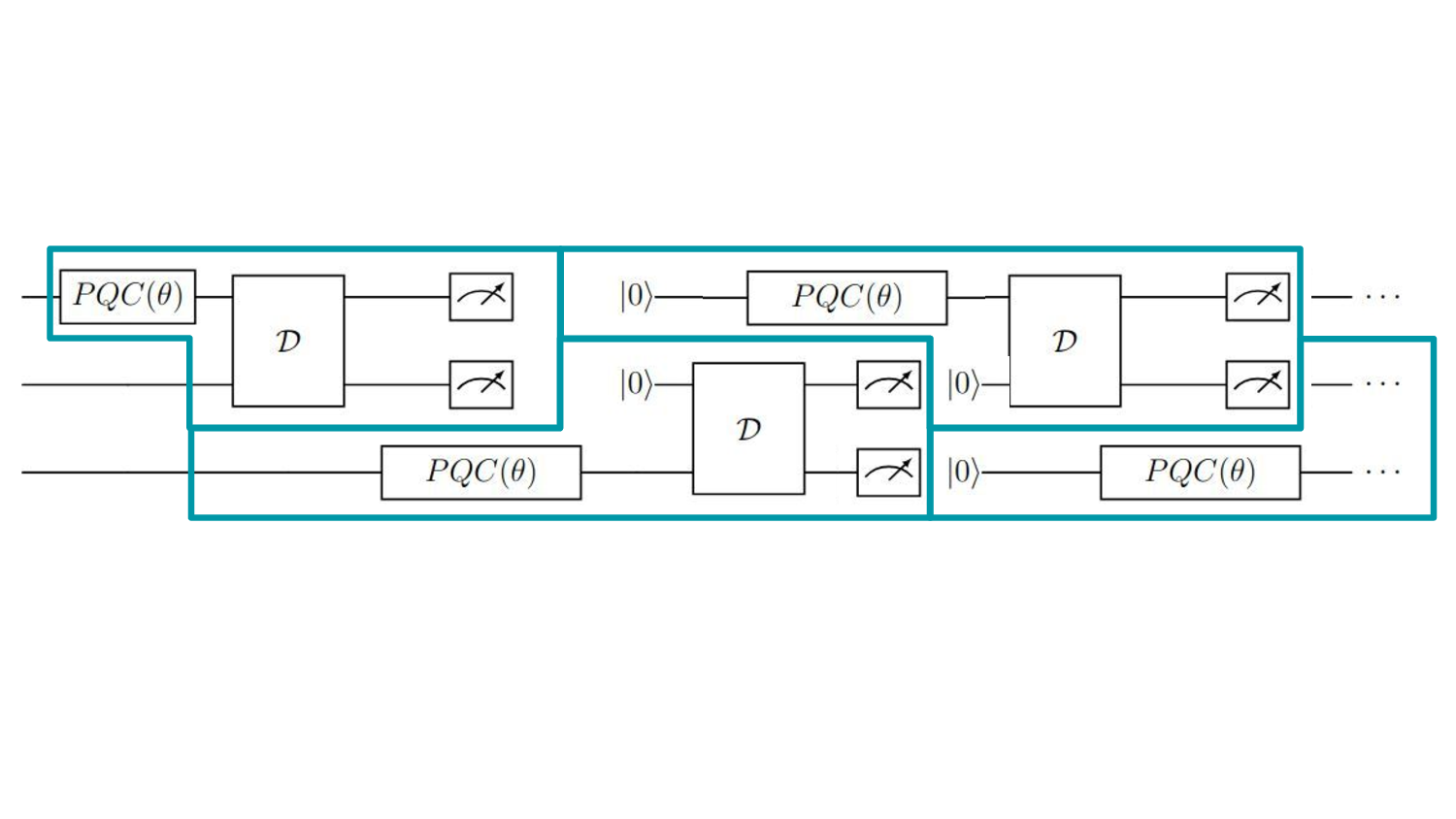}
        \caption{Possible pipelining structure for multiple VQE iterations using the compressed ansatz with decoder structure. For compactness, a 1-2 decompression is visualized but can be applied for other ratios such as 1-3, 2-4, etc.}
    \label{fig:pipeline}
\end{figure*}

We consider how to exploit the current quantum algorithm structure. One of the initial goals of this research was to reduce the number of qubits used by the VQE algorithm. In this work, we need to have a decoder that uses qubits equal to the number of qubits required in the original algorithm. This is needed to sample expectation values in order to find a minimal eigenvalue. While there is no overall reduction for the algorithm as a whole, we can exploit the temporary decrease in the number of qubits while being in the compressed state. In the context of using the calculation in an ab-initio molecular dynamics setting, we expect to do many energy calculations in rapid succession. We can pipeline these instructions to achieve better efficiency, which was not possible in the original algorithm. A visual illustration of such a pipelining effort is shown in Figure \ref{fig:pipeline}. Here, the resource-intensive decoding stage can be executed in parallel with the resource-efficient variational algorithm. This means we can run two algorithm executions (mostly) in parallel, using only $n+k$ qubits ($n$ encoder qubits, $k$ variational algorithm qubits). In the example of a 4-to-2 qubit encoding, only 6 qubits are needed instead of 8, with an extended runtime of one decoder depth.

Furthermore, the structure from Figure \ref{fig:pipeline} becomes especially interesting when the depth of the PQC is close to the depth of the decoder (with measurement and qubit reset). For this, both efforts of creating a more efficient scaling of the decoder and exploring the increasing PQC repetitions become especially relevant.

\subsection{Parameter training and neural network predictions}
In our use case, a lower number of parameters is required for the VQE optimization when compressing the PQC. In the case that the neural network predicted parameters are not sufficiently accurate, the prediction can still be used as a warm start for conventional optimization techniques as also suggested by Tao et al \cite{tao2022VQEPES}. This can significantly reduce optimization time. The reduction of parameters through compression also reduces the dimentionality of the solution space, further accelerating the optimization.

Finding parameters in an increasingly complex search space is considered one of the current prevalent challenges in variational quantum algorithms \cite{mcclean2018barren} (barren plateau). With our results, it could be explored how many similarities between different quantum problems are required to use a proxy or warm start for close convergence to the next problem. This is interesting in relatively similar atom configurations, but could possibly be extended to graph problems where the structure is roughly consistent.

This is especially relevant as we conclude that trainability through enforcing a gradual progression of parameters is not necessarily detrimental to the accuracy. Training with enforced gradients performs better on larger ansatzes, as opposed to starting from random parameters. This is likely due to initializing with parameters of a closely related structure, which effectively provides a warm start, allowing the optimizer to find an optimum faster. For ansatzes with more parameters, this is evidently beneficial for training.

A further point to touch upon is investigating how similar the circuit ansatz parameters remain when changing the atom configuration. In our rather simple case of the $H_2$ system, no drastic changes occur in changing the atomic distance. However, if we take the example of a phase change or excitation, the optimal circuit configuration could change as well. Further investigation is required as to how such mechanics will translate to the model predictions.

Beyond parameter optimization itself, improvements in molecular encoding passed to the neural network should be extended to more mature methods. To restrict the scope of this paper, the current encoding is only the atomic distance. However, far more sophisticated methods are generally used for classical MLP, ensuring e.g. rotational invariance \cite{pinheiro2021choosing}, of which the integration with VQE has been demonstrated before \cite{tao2022VQEPES}. This encoding can be included in the NN-AE-VQE framework to integrate this model into molecular dynamics.

\subsection{QAE challenges and opportunities}
One direct limitation of the QAE framework is the maximal compression rate that can be achieved with the QAE without significant losses is bounded by the problem Hamiltonian structure \cite{MA2023110659}. This was observed in the compression of the \ce{LiH} state, where compression from 8-3 qubits results in too much information loss for chemical accuracy.  Such limits need to be taken into account when evaluating how much of an advantage can be achieved. Here, better understanding of the quantum information structure is needed to predict where this limit lies.

Other QAE implementations have been presented in earlier literature \cite{bravo2021quantum, lamata2018quantum}, which can be considered when aiming to improve the presented implementation. Another structure to implement for scaling up is the quantum compression \cite{anand2022quantumcompressionclassicallysimulatable, achache2020denoisingquantumstatesquantum}. Here, a lower-order qubit compression ansatz is re-used, reducing the need to re-train parameters. For demonstration purposes this was not explicitly needed, but could be beneficial in practical application.

Choosing the correct ansatz for QAE compression has proven to be non-trivial. While the ansatz structure used for the \ce{H_2} molecule performed well in 4 to 2 qubit compression, this performance did not translate to the \ce{LiH} molecule using 8 qubits. Instead the hardware efficient SU2 structure performed significantly better. The challenge posed here, is that it is not yet well understood how to best select the ansatz structure for a given problem.

\subsection{Ansatz construction}
The ansatz selection problem extends further to the VQE calculations itself, as presented by the ansatz comparison in Table \ref{tab:LiH}. There is an intuition of why finding a good ansatz for the AE-VQE algorithm is challenging. First, as the AE-VQE method introduces a compressed space for VQE calculations, informed ansatz structures that make use of e.g. symmetry \cite{meyer2023exploiting} cannot be exploited as easily. Nor is it feasible to construct  physically motivated methods, as the state preparation is no longer organized as a physically structured state. Although it might retain this property in higher dimensionality, there is no way to extract this structure effectively as it changes with each QAE instance.

Another important element we would like to consider is the effect of increasing the number of (compressed) ansatz layers. In the current implementation, using a single layer proved to be effective for the \ce{H_2} molecule and three for the \ce{LiH} molecule. As shown by Bravo-Prieto et al., the number of layers can find a critical point at which the error rate is significantly reduced \cite{bravo2020scaling}. Increasing the number of layers in the compressed space is favorable in terms of circuit depth compared to an uncompressed ansatz, therefore, this could be an advantage to exploit. Interestingly, it appears that in the AE-VQE structure this critical point was found at three layers, whereas for the UCCSD-VQE this lies at four layers.

This touches upon the critical aspect of choosing the right ansatz. Sim et al. have demonstrated that for certain ansatz selections, the expressivity saturates \cite{sim2019expressibility}. This indicates that not all ansatz designs are suited for layering. Notably, the expressivity is also not a direct indication of VQE performance. As stated before, further research is required to systematically choose the right ansatz depending on the problem.

The work by Bravo-Prieto et al. has shown that increasing PQC layers increases the accuracy of a VQE calculation after a certain threshold of number of layers has been reached \cite{bravo2020scaling}. Going by the aforementioned metrics, the assumption is that by increasing the number of layers of a PQC, the expressivity is increased to the point that the optimal solution can be found. Yet, this is not straightforward. Numerical results by Sim et al. show that while for some hardware-efficient PQC this is the case, several PQC structures exhibit a limit to the expressivity that can be achieved \cite{sim2019expressibility}. Furthermore, the role of the entangling structure is not yet fully explored in this context. Adding to this, by increasing the number of layers, the effect of barren plateaus increases. This, in turn, imposes challenges on trainability.

In our work we identify a circuit optimization in terms of gates and parameters. By examining the parameters of optimized instances, we identify gates that stay constant for the same molecule for different configurations. This would allow for replacing parameterized gates with constant gates, which further saves neural network training for the parameters. Furthermore, for larger systems, the number of parameters of the circuit ansatz will be more significant and slowing down VQE calculations (or training of neural networks for predictions), giving rise to the need to optimize in terms of circuit parameters.

One possible approach for optimizing the circuit ansatz is to do quantum architecture search \cite{bilkis2023semi, du2022quantum, kuo2021quantum}. This method allows for the circuit to be optimized in terms of gates. Through optimization methods gates can be added or removed, to find a minimal configuration for the specific problem instance. While this gives significant overhead to an algorithm, this is a more structured approach compared to the trial and error that is currently prevalent in QML.

With these concerns, a fundamental knowledge gap is identified in the QML field. For unstructured quantum data, there is no efficient approach to constructing a problem-specific PQC. Crucial understandings of QML problems must be further explored, such as defining and determining the expressivity and entangling capacity of PQC, required for a specific problem. As such, we believe that efforts towards this target will be crucial for advancing the field of QML.

\section{Conclusion}
\label{sec:conclusion}

This work presents a novel approach to use a quantum auto-encoder to reduce the resources required for the VQE ansatz. Additionally, we provide a structured approach to prepare training data of the ansatz parameters for effective learning by a classical neural network. Through these advances, we present progression toward a quantum utility for high-accuracy MD simulations despite current hardware constraints.
We demonstrate the validity of the proof-of-concept model, by testing the model on the \ce{H_2} molecule as a minimal working example. Furthermore, we extend our model to a \ce{LiH} model, reducing quantum circuit length by $98.3 \%$ without loss of accuracy and identify fundamental QML challenges. For both systems, we demonstrate an error rate well below chemical accuracy.
With the proposed structure, we introduce a novel approach to construct quantum variational circuits and introduce the exploitation of their unique advantages. Our methodology integrates VQE calculations more efficiently with molecular dynamics (MD) simulations, paving the way for significant improvements compared to existing VQE implementations.
By combining VQE with machine learning, we demonstrate a more efficient trade-off of quantum resource requirements and accuracy. Reducing quantum resources will allow for larger molecular systems to be simulated while retaining chemical accuracy. This is critical for high-throughput screening of novel materials such as battery and high-entropy materials. 
 We provide the code to run all experiments on Github: \url{https://github.com/koenmesman/NN-AE-VQE}.
Moreover, we identify a critical knowledge gap on how to structure PQC for unstructured data. We suggest a path to overcome this challenge, and we believe that advancing towards more structured approaches to ansatz generation will be crucial for the further development of quantum machine learning. 
By reducing quantum resource requirements through compression and structured parameter learning, this work lays the foundation for scaling hybrid quantum algorithms to larger and more complex systems. In doing so, it moves a step closer to integrating quantum simulations into practical materials discovery and molecular dynamics workflows.

\bibliography{bibliography}

\end{document}